\documentclass[11pt,a4paper,twoside,reqno, draft]{amsart}
\usepackage{amsfonts,amssymb,euscript,amscd}
\DeclareFontFamily{OML}{cyr}{} \DeclareFontShape{OML}{cyr}{m}{n}{
      <5> <6> <7> <8> <9> gen * wncyr
      <10> <10.95> <12> <14.4> <17.28> <20.74> <24.88> wncyr10
      }{}
\DeclareSymbolFont{rusletters}{OML}{cyr}{m}{n}
\DeclareSymbolFontAlphabet{\rusmath}{rusletters}
\DeclareMathSymbol\re{\rusmath}{rusletters}{"03}
\newtheorem{theorem}{Theorem}[section]
\newtheorem{proposition}[theorem]{Proposition}
\newtheorem{lemma}[theorem]{Lemma}
\newtheorem{corollary}[theorem]{Corollary}
\newtheorem{remark}[theorem]{Remark}
\newtheorem{example}[theorem]{Example}

\newcommand*{\R}{\mathbb R} 
\newcommand*{\im}{\mathop{\rm Im}\nolimits} 
\newcommand*{\tr}{\mathop{\rm tr}\nolimits} 
\newcommand*{\id}{\mathop{\rm id}\nolimits} 
\newcommand*{\pr}{\mathop{\rm pr}\nolimits} 
\newcommand*{\E}{\EuScript E} 
\newcommand*{\Crv}{\EuScript R} 
\newcommand*{\C}{\EuScript C} 
\newcommand*{\Cn}{\EuScript V} 
\newcommand*{\RP}{\mathbb R\mathrm P} 
\newcommand{\kch}{\mathbin{\rule{5pt}{0.5pt}\rule{0.5pt}{7pt}}} 
\newcommand*{\Smbl}{\mathop{\rm Smbl}\nolimits} 
\newcommand*{\Q}{\EuScript Q} 
\newcommand*{\D}{\EuScript D} 
\newcommand*{\Orb}{\mathop{\rm Orb}\nolimits} 
\newcommand*{\codim}{\mathop{\rm codim}\nolimits} %
\newcommand*{\Prj}{\EuScript P} 
\newcommand*{\X}{\EuScript X} 
\title[Monge--Amp\`ere equations]{Differential invariants of
  generic hyperbolic Monge--Amp\`ere equations}
\author{Michal Marvan}
\address{Mathematical Institute of the Silesian University in
Opava, Na Rybn\'{\i}\v{c}ku~1, 746 01 Opava, Czech Republic}
\email{Michal.Marvan@math.slu.cz}
\author{Alexandre M. Vinogradov}
\address{University of Salerno,
Dipartimento di Matematica ed Informatica, Facolta' di Scienze
Matematiche, Fisiche e Naturali, Universita' degli Studi di
Salerno, Via Ponte del Melillo, 84084 Fisciano (SA), Italy,
and Istituto Italiano di Fisica Nucleare, Italy}
\email{vinograd@unisa.it}
\author{Valery A. Yumaguzhin}
\address{Program Systems Institute of RAS,
    152020, Pereslavl'-Zalesskiy, m. Botik, Russia}
\email{yuma@diffiety.botik.ru}
\date{7 April 2006}

\begin{document}
\begin{abstract} In this paper basic differential
  invariants of generic hyperbolic Monge--Amp\`ere equations with
  respect to contact transformations are constructed and the
  equivalence problem for these equations is solved.
\end{abstract}

\maketitle

\section{Introduction}
With this paper we start a systematic study of differential
invariants of Monge--Amp\`ere equations, with the objective of the 
classification problem, methods of integration and other applications. 
Complete proofs of the results announced in ~\cite{MVY} are presented.
We are interested in the classical case of two independent variables. 
The Monge--Amp\`ere equations equations merit a special attention due 
to a large spectrum of various applications, first of all, in 
differential geometry and mathematical physics. Moreover, they form a 
natural testing area for new methods emerging in the modern theory of 
nonlinear PDE's. 

In spite of more than 200 years of history of Monge--Amp\`ere
equations and numerous publications devoted to them it would be an
exaggeration to say that their nature is well understood. An
important success was establishing the existence and uniqueness
theorems by Lewy and others (see \cite{Lewy,H-W} for local aspects
and \cite{Tuni} for global ones). The classical Monge integration
method was modernized by Matsuda \cite{Mats1,Mats2} and Morimoto
\cite{Mori}, etc. Our interest in differential invariants is
motivated not only by the classification problem but, no less, by
hopes that they could illuminate many aspects of the theory of
Monge--Amp\`ere equations.

According to \cite{V1} (see also \cite{ALV}) scalar differential
invariants provide a key to solving the classification problem for
any kind of geometrical structures. In fact, geometrical
structures of a given type are classified by solutions of a
naturally associated \emph{classifying} (differential) equation,
which describes ``family ties'' connecting the corresponding scalar
differential invariants. More exactly, scalar differential
invariants are smooth functions on the \emph{classifying diffiety},
which is the infinite prolongation of the classifying equation.
This diffiety has, generally, singularities and its singular
strata classify those geometrical structures that possess
nontrivial symmetries. Each of these strata is also an infinitely
prolonged differential equation in a lesser number of independent
variables. For instance, homogeneous structures correspond to the
zero-dimensional case. So, the classification problem consists of
a complete description of all strata composing the classifying
diffiety and, therefore, involves a complete symmetry analysis of
the geometric structures under consideration. The interested
reader will find an illustration of the above said in \cite{Vi-Yu}
where plane 3-webs, a rather simple geometrical structure, is
considered.

The classification problem for Monge--Amp\`ere equations dates
back to Sophus Lie.
%
%
For modern proofs of Lie's theorems, classification  problems for
various strata of Monge-Amp\`ere equations see, e.g.,
\cite{LRC,Kr1,Kr2,Kr3,Ku,Tch} and references therein.
%
%
In this paper we interpret a hyperbolic Monge--Amp\`ere equation
as a pair of 2-di\-men\-si\-o\-nal, skew-othogonal non-lagrangian
subdistributions of the contact distribution on 5-dimensional
contact manifolds. Another approach to these equations was
developed by V.V. Lychagin in \cite{Ly1,Ly2}. We look for
differential invariants of Monge--Amp\`ere equations, not only
scalar, with respect to the group of contact transformations. Here
we limit ourself to the case of generic hyperbolic equations. This
is motivated by two reasons. First, the study of singular strata
benefits much from the knowledge of the generic one. Second, for
the hyperbolic equations differential invariants are easier
visible due to the existence of bicharacteristics.
%

Differential invariants found in this paper give a solution of the
classification problem for generic hyperbolic equations. This
solution requires a substantial computer support in analysis of
concrete cases and a further work is necessary to improve its
efficiency.

Differential invariants for elliptic and parabolic Monge--Amp\`ere
equations can be obtained more or less straightforwardly by
following the approach developed in this paper. This and a the study
of singular strata will be the subject of subsequent publications.

\section{Preliminaries}

Below, all manifolds and maps are supposed to be smooth. By
$[f]_p^k\,,\;k=0,1,2,\ldots,\infty,$ we denote the $k$-jet of a
map $f$ at a point $p$. $\R$ stands for the field of real numbers,
and $\R^n$ for the $n$-dimensional arithmetic space.

\subsection{Jet bundles}
Here we recall necessary definitions and facts about jet bundles,
see~\cite{KLV,KV}.

Let $M$ be an $n$-dimensional manifold,  $E$  an $n+m$-dimensional
manifold and
$$
      \pi:E\longrightarrow M\,.
$$
a fiber bundle. By
$$
  \pi_k:J^k\pi\to M\,,\quad\pi_k:[S]^k_p\mapsto p\,,
  \quad k=0,1,2,\ldots
$$
we denote the bundle of all $k$-jets of sections of $\pi$.
For any $l > m \geq 0$, the natural projection is defined as
$$
  \pi_{l,m}:J^l\pi\to J^m\pi\,,\quad
  \pi_{l,m}:[S]^l_p\mapsto [S]^m_p\,.
$$
Any section $S$ of $\pi$ generates the section $j_kS$ of the
bundle $\pi_k$ by the formula
$$
  j_kS : p\mapsto [S]^k_p\,.
$$
Put
$$
  L^k_S=\im j_kS\,.
$$
Let $\theta_{k+1}$ be an arbitrary point of $J^{k+1}\pi$,
$\theta_k=\pi_{k+1,k}(\theta_{k+1})$, and $T_{\theta_k}(J^k\pi)$
the tangent space to $J^k\pi$ at the point $\theta_k$. Then
$\theta_{k+1}$ defines the subspace $K_{\theta_{k+1}}\subset
T_{\theta_k}(J^k\pi)$ by the formula
$$
  K_{\theta_{k+1}}=T_{\theta_k}(L^k_S)\,.
$$
Clearly, $\theta_{k+1}$ is identified with $K_{\theta_{k+1}}$.
It is easy to prove that
\begin{equation}\label{DrSmDc}
  T_{\theta_k}(J^k\pi)=K_{\theta_{k+1}}\oplus
  T_{\theta_k}(\pi_k^{-1}(p))\,.
\end{equation}
Consider all submanifolds of the form $L^k_S$ containing
$\theta_k$. Subspace spanned by their tangent spaces
$T_{\theta_k}(L^k_S)$ is denoted by $\C(\theta_k)$ and
it is called the {\it Cartan plane at $\theta_k$}. The
distribution
$$
  \C_k:\theta_k\mapsto \C(\theta_k)
$$
is called the {\it Cartan distribution on $J^k\pi$}. The
distribution $\C_k$, $k\geq 1$, can be defined as the kernel of
the {\it Cartan form}
$$
  U_k=\pr_2\circ\,(\pi_{k,k-1})_*\,,
$$
where $\pr_2:T_{\theta_{k-1}}(J^{k-1}\pi)\to
T_{\theta_{k-1}}(\pi_{k-1}^{-1}(p))$ is the projection generated
by direct sum decomposition \eqref{DrSmDc}.
\subsection{The contact structure}
%


Consider the trivial bundle
$$
      \tau:\R^2\times\R\longrightarrow\R^2\,,\quad
      \tau:(\,x, y, z\,)\mapsto (\,x, y\,)\,.
$$
By $x,y,z$, $p = z_x$, $q = z_y$, $r = z_{xx}$, $s = z_{xy}$,
$t = z_{yy}$ we denote the standard coordinates in $J^2\tau$.

The Cartan distribution $C_1$ on $J^1\tau$ is identical to the
{\it contact structure} on $J^1\tau$.
The corresponding contact 1-form $U_1$ has the canonical form
$$
  U_1 = dz - p\,dx - q\,dy\,.
$$
in the standard coordinates.

A diffeomorphism $\varphi:J^1\tau\to J^1\tau$ is called a {\it contact
transformation} if it preserves the Cartan distribution. Obviously,
a diffeomorphism $\varphi$ is a contact transformation iff there exist a
nowhere vanishing function $\lambda$ such that
$$
 \varphi^*(U_1)=\lambda\,U_1\,.
$$

Any contact transformation $\varphi$ can be lifted to the diffeomorphism
$$
  \varphi^{(1)}_{\tau}:J^2\tau\longrightarrow J^2\tau
$$
by the formula
$$
  \varphi^{(1)}_{\tau}: \theta_2\equiv K_{\theta_2}\mapsto
  \varphi_*(K_{\theta_2})\equiv\tilde\theta_2
  =\varphi^{(1)}_{\tau}(\theta_2)\,.
$$
If $\varphi$ is defined on an open set $V\subset
J^1\tau$, then $\varphi^{(1)}_{\tau}$ is defined on
an open, everywhere dense subset of $\tau_{2,1}^{-1}(V)$.

A vector field $Z$ in $J^1\tau$ is a {\it contact vector field} if
its flow $\varphi_t$ consists of contact transformations. Clearly,
$Z$ is a contact vector field iff there exist a function $\lambda$
such that
$$
 L_Z(U_1)=\lambda\,U_1\,,
$$
where $L_Z$ is the Lie derivative with respect to $Z$.

There exists a natural one-to-one correspondence between the set
of all contact vector fields in $J^1\tau$ and the set of all
functions in $J^1\tau$. It is defined by the formula
$$
  Z\mapsto f=Z\kch U_1\,.
$$
The function $f=Z\kch U_1$ is called the {\it generating function
of the contact vector field $Z$}. The contact vector field $Z$
corresponding to $f$ is denoted by $Z_f$. In standard coordinates,
the field $Z_f$ is given by the formula
\begin{equation}\label{CntVctFld}
  Z_f=-f_p\frac{\partial}{\partial x}
                 -f_q\frac{\partial}{\partial y}
                 +(f-pf_p-qf_q)\frac{\partial}{\partial z}
                 +(f_x+pf_z)\frac{\partial}{\partial p}
                 +(f_y+qf_z)\frac{\partial}{\partial q}.
\end{equation}

\subsection{Operations over vector-valued forms}
Let $M$ be a smooth $n$-di\-men\-si\-o\-nal manifold,
$\Lambda^i(M)$ the $C^{\infty}(M)$-module of $i$-forms on $M$ and
$D(M)$ the $C^{\infty}(M)$-module of vector fields on $M$. Let
$\alpha\in\Lambda^k(M)$, $\beta\in\Lambda^r(M)$, and $X, Y\in
D(M)$. Then the Fr\"olicher--Nijenhuis bracket $[\![\cdot\,,
\cdot]\!]$ of the vector-valued forms $\alpha\otimes X$ and
$\beta\otimes Y$ is defined by the formula
\begin{multline*}
     [\![\alpha\otimes X,\,\beta\otimes Y]\!]\\
     =\alpha\wedge\beta\otimes\bigl[X,Y\bigr]
     +\alpha\wedge X(\beta)\otimes Y
     -Y(\alpha)\wedge\beta\otimes X\\
     +(-1)^kd\alpha\wedge (X\kch\beta)\otimes Y
     -(-1)^k(Y\kch\alpha)\wedge d\beta\otimes X\,,
\end{multline*}
see \cite{FN}.

The contraction $\,\kch$ of forms $\alpha\otimes X$
and $\beta\otimes Y$ is defined by the formula
$$
     (\alpha\otimes X)\kch(\beta\otimes Y)
     =\alpha\wedge (X\kch\beta)\otimes Y\,.
$$

\subsection{Projectors and their curvatures}
The following simple construction allows one to associate a vector
valued 2-form with a projector. Namely, let $P,Q \in D(M)$ be
endomorphisms of the $C^{\infty}(M)$-module $D(M)$ such that
$QP=0$. Then
\begin{equation}
\label{Omega}
\Omega_{Q,P}(X,Y)=Q[P(X),P(Y)], \quad X,Y\in D(M),
\end{equation}
obviously, is skew-symmetric and $C^{\infty}(M)$-bilinear, i.e., a
vector valued form. More precisely, it takes values in
$\im\;Q\subset D(M)$. If $P:D(M)\rightarrow D(M)$ is a projector,
i.e., $P^2=P$, then the associated
\emph{curvature form} of $P$ is defined to be
\begin{equation}
\label{Crv}
\Crv_P=\Omega_{I-P,P}
\end{equation}
with $I=\id_{D(M)}$.

\section{Hyperbolic Monge--Amp\`ere equations}
\subsection{Monge--Amp\`ere equations}
The Monge--Amp\`ere equation is a partial differential equation of
the form
\begin{equation}\label{MAeq}
      N(z_{xx}z_{yy}-z_{xy}^2)+Az_{xx}+Bz_{xy}+Cz_{yy}+D=0\,,
\end{equation}
where $x$, $y$ are independent variables, $z$ is a dependent
variable, $z_{xx}=\partial^2z/\partial x^2$,
$z_{xy}=\partial^2z/\partial x\,\partial y$,
$z_{yy}=\partial^2z/\partial y^2$, and coefficients $N$,
$A$, $B$, $C$, $D$ are functions of $x$, $y$, $z$,
$z_x = \partial z/\partial x$ and $z_y = \partial z/\partial y$.

We identify equation \eqref{MAeq} with the submanifold
$\E$ of the jet bundle $J^2\tau$ determined by the equation
\begin{equation}\label{MAeq1}
     N(rt - s^2) + Ar + Bs + Ct + D = 0\,.
\end{equation}
Obviously,
$$
  \tau_{2,1}(\E)=J^1\tau\,.
$$

Let $\theta_2\in\E$, $\tau_{2,1}(\theta_2)=\theta_1$, and
$F_{\theta_1}$ be the fiber of the projection $\tau_{2,1}$ over
the point $\theta_1 \in J^1\tau$. Then the subspace
$$
  \Smbl_{\theta_2}\E=T_{\theta_2}\E\cap T_{\theta_2}F_{\theta_1},
$$
where $T_{\theta_2}\E$ is the tangent space to $\E$ at $\theta_2$
is called the {\it symbol of the equation $\E$ at the point}
$\theta_2\in\E$. In terms of standard coordinates, $\Smbl_{\theta_2}\E$
is described by the linear equation
\begin{equation}\label{SmblMAeq}
     N (t\tilde r + r\tilde t - 2s\tilde s)
      + A\tilde r + B\tilde s + C\tilde t = 0,
\end{equation}
where $\tilde r$, $\tilde s$, $\tilde t$ are the standard
coordinates in $T_{\theta_2}F_{\theta_1}$ generated by the
standard coordinates on $J^2\tau$.

A point $\theta_2\in\E$ can be elliptic, parabolic, or hyperbolic.
To introduce these notions, let us consider a one-dimensional subspace
$P\subset \C(\theta_1)$ such that $(\tau_1)_*P\ne 0$.
By definition, put
$$
     l(P)=\{\,\theta_2\in F_{\theta_1}\,\bigr|\,
            P\subset K_{\theta_2}\,\}\,.
$$
The submanifold $l(P)$ is called a {\it 1-ray}.
In terms of standard coordinates, let
$\theta_1=(x, y, z,p, q)$, $P=\langle v\rangle$ and
\begin{equation}\label{Vctr}
     v=\zeta_1\frac{\partial}{\partial x}
       +\zeta_2\frac{\partial}{\partial y}
       +\mu\frac{\partial}{\partial z}
       +\eta_1\frac{\partial}{\partial p}
       +\eta_2\frac{\partial}{\partial q}\,.
\end{equation}
Then $(\tau_1)_*P\ne 0$ means that
\begin{equation}\label{NnSnglr}
  (\zeta_1,\zeta_2) \neq (0,0)\,,
\end{equation}
$v\in\C(\theta_1)$ means that
\begin{equation}\label{InC}
     \mu=\zeta_1p+\zeta_2q\,,
\end{equation}
and $P\subset K_{\theta_2}$ means that
\begin{equation}\label{InR}
\left\{
     \begin{aligned}
     \eta_1 &= \zeta_1r+\zeta_2s\,,\\
     \eta_2 &= \zeta_1s+\zeta_2t\,,
     \end{aligned}\right.
\end{equation}
where $r, s, t$ are the standard coordinates of $\theta_2$ in the
fiber $F_{\theta_1}$.
From system \eqref{InR}, we see that $l(P)$ is an affine straight
line in $F_{\theta_1}$.
By $\ell_{\theta_2}(P)$ we denote the tangent space
$T_{\theta_2} l(P)$ to $l(P)$ at the point $\theta_2\in l(P)$.
We call it a {\it 1-ray subspace}.
In terms of the standard coordinates $\tilde r, \tilde s, \tilde t$
in $T_{\theta_2} F_{\theta_1}$, vectors of $\ell_{\theta_2}(P)$
satisfy
\begin{equation}\label{1_rSp}
   \left\{
     \begin{aligned}
       \zeta_1\tilde r+\zeta_2\tilde s&=0\,,\\
       \zeta_1\tilde s+\zeta_2\tilde t&=0\,,
     \end{aligned}\right.
\end{equation}
Obviously, $\ell_{\theta_2}(P)$ is spanned by the vector
\begin{equation}
\label{RaySp}
      (\,\tilde r,\,\tilde s,\,\tilde t\,) =
       (\,\zeta_2^2,\, -\zeta_1\zeta_2,\,\zeta_1^2\,)\,.
\end{equation}
Taking into account \eqref{NnSnglr}, we observe that all 1-ray subspaces
form the cone
$$
     \Cn_{\theta_2} = \{\,\tilde r \tilde t- \tilde s^2 = 0\,\}
$$
in the tangent space $T_{\theta_2}F_{\theta_1}$.
This cone is called the {\it cone of singular square forms}.
Obviously, the intersection
$\Smbl_{\theta_2}\E\cap\,\Cn_{\theta_2}$ is either zero, or a
single 1-ray subspace, or two 1-ray subspaces. Correspondingly,
the point $\theta_2\in\E$ is then called {\it elliptic}, {\it
parabolic} or {\it hyperbolic}.  It is not difficult to prove that
a contact transformation takes an elliptic, parabolic, or
hyperbolic point to an elliptic, parabolic, or hyperbolic point,
respectively. The equation $\E$ is called {\it elliptic}, {\it
parabolic} or {\it hyperbolic} if all its points are elliptic,
parabolic or hyperbolic, respectively. In this work, we consider
hyperbolic Monge--Amp\`ere equations only. It is easy to see that
$\E$ is hyperbolic iff its coefficients satisfy the condition
\begin{equation}\label{Hprbl}
    \Delta = B^2 - 4AC + 4ND > 0\,.
\end{equation}


\subsection{Skew-orthogonal distributions}
Following \cite{V2}, we show that a hyperbolic
Monge--Amp\`ere equation is equivalent to a pair of
skew-or\-tho\-go\-nal two-dimensional distributions in the Cartan
distribution on $J^1\tau$.

Let $\theta_1$ be an arbitrary point of $J^1\tau$.
By $\Q_{\theta_1}$ we denote the union of all one-dimensional
subspaces $P$ of $\C(\theta_1)$ such that $\tau_* P \ne 0$
and the 1-ray $l(P)$ is tangent to $\E$ at least at one point.
\begin{proposition}\label{BrkEqtn}
Let $\E$ be a hyperbolic Monge--Amp\`ere equation.
Then $\Q_{\theta_1}$ is the union of two-dimensional subspaces
$\D^1_{\E}(\theta_1)$ and $\D^2_{\E}(\theta_1)$
of the Cartan plane $\C(\theta_1)$, so that
\begin{enumerate}
\item $\C(\theta_1)
    = \D^1_{\E}(\theta_1)\oplus\D^2_{\E}(\theta_1)$,
\item $\D^1_{\E}(\theta_1)$ and
         $\D^2_{\E}(\theta_1)$ are skew-orthogonal with respect
to the symplectic form $dU_1 = dx \wedge dp + dy \wedge dq$
on~$\C$.
\end{enumerate}
\end{proposition}
\begin{proof} We prove this proposition for Monge--Amp\`ere
equations such that $N\neq 0$. The proof for $N=0$ follows from
the fact that every Monge--Amp\`ere equation can be transformed to
one with $N\neq 0$ by an appropriate contact transformation.

Let $v\in\Q_{\theta_1}$ and $P=\langle v\rangle$. The condition
for $l(P)$ to be tangent to $\E$ can be written in the following
way. We can assume that $v$ is of the form \eqref{Vctr}. Then the
vector of fiber coordinates $(\zeta_2^2,
-\zeta_1\zeta_2,\zeta_1^2)$ is tangent to $l(P)$. Now using
\eqref{SmblMAeq} we deduce that $l(P)$ is tangent to $\E$ iff
$$
  N(r\zeta_1^2+2s\zeta_1\zeta_2+t\zeta_2^2)
  +A\zeta_2^2-B\zeta_1\zeta_2+C\zeta_1^2=0\,.
$$
Taking into account that the coordinates $\zeta_i$ and $\eta_i$ of
$v$ are connected by equations \eqref{InR}, we reduce this
equation to the form
\begin{equation}\label{TngntFbrE}
    N(\zeta_1\eta_1+\zeta_2\eta_2)
    +A\zeta_2^2-B\zeta_1\zeta_2+C\zeta_1^2=0\,.
\end{equation}

Then in view of \eqref{NnSnglr} we assume that $\zeta_1\neq 0$
(the case $\zeta_2\neq 0$ is analogous). Then from \eqref{InR} we
get
$$
  r=\frac{1}{\zeta_1^2}(\eta_1\zeta_1-\eta_2\zeta_2
        +\zeta_2^2t)\,,\quad
  s=\frac{1}{\zeta_1}(\eta_2-\zeta_2t)\,.
$$
Substituting these expressions for $r$ and $s$ in equation
\eqref{MAeq1} and taking into account equation \eqref{TngntFbrE},
we obtain the equation
\begin{equation}\label{MAeq2}
  N\eta_2^2+(A\zeta_2-B\zeta_1)\eta_2-A\zeta_1\eta_1-D\zeta_1^2=0.
\end{equation}

Solving the system of equations \eqref{TngntFbrE} and
\eqref{MAeq2} with respect to $\eta_1$ and $\eta_2$, we find
$$
  \eta_1=\frac{(B\mp\sqrt\Delta)\zeta_2-2C\zeta_1}{2N}\,,\quad
  \eta_2=\frac{(B\pm\sqrt\Delta)\zeta_1-2A\zeta_2}{2N}\,.
$$
Finally, in view of \eqref{InC}, we see that
\begin{multline}
\label{vinD}
    v = \zeta_1\biggl(\frac{\partial}{\partial x}
      + p\frac{\partial}{\partial z}
      - \frac{C}{N}\frac{\partial}{\partial p}
      + \frac{B\pm\sqrt\Delta}{2N}\frac{\partial}{\partial q}
        \biggr)\\
      + \zeta_2\biggl(\frac{\partial}{\partial y}
      + q\frac{\partial}{\partial z}
      + \frac{B\mp\sqrt\Delta}{2N}\frac{\partial}{\partial p}
      - \frac{A}{N}\frac{\partial}{\partial q}\biggr).
\end{multline}
This proves that $\Q_{\theta_1} =\langle X_1,X_2\rangle\cup\langle
X_3,X_4\rangle$ with
\begin{equation}
\begin{aligned}\label{Dstr}
     X_1 &= \frac{\partial}{\partial x}
         + p\frac{\partial}{\partial z}
         - \frac{C}{N}\frac{\partial}{\partial p}
         + \frac{B-\sqrt\Delta}{2N}\frac{\partial}{\partial q}\,,\\
     X_2 &= \frac{\partial}{\partial y}
         + q\frac{\partial}{\partial z}
         + \frac{B+\sqrt\Delta}{2N}\frac{\partial}{\partial p}
         - \frac{A}{N}\frac{\partial}{\partial q},\\
     X_3 &= \frac{\partial}{\partial x}
         + p\frac{\partial}{\partial z}
         - \frac{C}{N}\frac{\partial}{\partial p}
         + \frac{B+\sqrt\Delta}{2N}\frac{\partial}{\partial q}\,,\\
     X_4 &= \frac{\partial}{\partial y}
         + q\frac{\partial}{\partial z}
         + \frac{B-\sqrt\Delta}{2N}\frac{\partial}{\partial p}
         - \frac{A}{N}\frac{\partial}{\partial q}\,.
\end{aligned}
\end{equation}
Put
$$
  \D^1_{\E}(\theta_1)=\langle X_1,X_2\rangle\,,\quad
  \D^2_{\E}(\theta_1)=\langle X_3,X_4\rangle\,.
$$
Now it is straightforward  to verify that subspaces
$\D^1_{\E}(\theta_1)$ and $\D^2_{\E}(\theta_1)$ are
skew-orthogonal and
$\D^1_{\E}(\theta_1)\cap\D^2_{\E}(\theta_1)=\{0\}$. This completes
the proof.
\end{proof}

From \eqref{Dstr} we see that for a Monge--Amp\`ere equation such
that $N\neq 0$, the map $\tau_{1*}$ projects
$\D^1_{\E}(\theta_1)$ and $\D^2_{\E}(\theta_1)$ onto the tangent
space to the base of the bundle $\tau$ without degeneration.

It should be noted that if $N=0$ (that is, if $\E$ is a quasilinear
second order PDE), then the projections
$\tau_{1*}\bigl(\D^1_{\E}(\theta_1)\bigr)$ and
$\tau_{1*}\bigl(\D^2_{\E}(\theta_1)\bigr)$ are one-dimensional.

Thus an arbitrary hyperbolic Monge--Amp\`ere equation generates two
2-dimensional skew-orthogonal subdistributions of the Cartan
distribution $\C_1$ in $J^1\tau$.
\begin{proposition}\label{reconstr} Let $\E$ be a hyperbolic Monge--Amp\`ere
     equation. Then $\theta_2 \in \E$ if and only if one of the
     following equivalent conditions holds:
     \begin{enumerate}
     \item $K_{\theta_2}\cap\D^1_\E(\theta_1)$ is 1-dimensional,
     \item $K_{\theta_2}\cap\D^2_\E(\theta_1)$ is 1-dimensional.
     \end{enumerate}
\end{proposition}
\begin{proof} As in the proof of Proposition
\ref{BrkEqtn} one can assume that $N\neq 0$.

Let $\theta_2\in\E$. Then
$\Smbl_{\theta_2}\E\cap\Cn_{\theta_2}=\ell_{\theta_2}(\langle
v\rangle)\cup\ell_{\theta_2}(\langle \tilde v \rangle)$, where
$\ell_{\theta_2}(\langle v \rangle)$ and
$\ell_{\theta_2}(\langle\tilde v \rangle)$ are different straight
lines and, so, vectors $v$ and $\tilde v$ are independent. They
are skew-orthogonal, since $K_{\theta_2}$ is a Lagrangian plane in
$\C(\theta_1)$ and, by definition of $\Q_{\theta_1}$, $v,\,\tilde
v\in\Q_{\theta_1}$. This means that $K_{\theta_2}$ intersects
planes $\D^1_\E(\theta_1)$ and $\D^2_\E(\theta_1)$ along $\langle
v\rangle$ and  $\langle\tilde v\rangle$, respectively.

Let $\theta_2$ be a point of $J^2\tau$ such that $K_{\theta_2}$
intersects the plane $\D^1_\E(\theta_1)$ along a straight line,
that is, $K_{\theta_2} \cap \D^1_\E(\theta_1) = \langle v\rangle$.
By substituting coordinates $\eta_1,\eta_2$ of the vector $v$
given by formula \eqref{vinD} into eq. \eqref{InR}, we obtain
\begin{eqnarray*}
\biggl(r + \frac CN\biggr) \zeta_1
    + \biggl(s - \frac {B - \sqrt\Delta}{2N}\biggr) \zeta_2 &=& 0,
\\
\biggl(s - \frac {B + \sqrt\Delta}{2N}\biggr) \zeta_1
    + \biggl(r + \frac AN\biggr) \zeta_2 &=& 0.
\end{eqnarray*}
By hypothesis this system is of rank 1 (cf. \eqref{NnSnglr}) and
hence its determinant is zero. Now it remains to note that this is
exactly equation~\eqref{MAeq1} and, so, $\theta_2\in\E$. The case
of $\D^2_\E(\theta_1)$ differs only by the sign at $\sqrt\Delta$.
\end{proof}
An important consequence of this proposition is that a hyperbolic
Monge--Amp\`ere equation $\E$ is completely determined by one of
the associated distributions $\D^i_\E$, $i=1,2$.

Thus, every hyperbolic Monge--Amp\`ere equation $\E$ is naturally
equivalent to a pair of 2-dimensional, skew-othogonal
non-lagrangian subdistributions $\D^1_\E$, $\D^2_\E$ of the Cartan
distribution $\C_1$ in $J^1\tau$. In particular, the equivalence
problem for hyperbolic Monge--Amp\`ere equations with respect to
contact transformations may be interpreted as the equivalence
problem for pairs of 2-di\-men\-si\-o\-nal, skew-orthogonal
non-lagrangian subdistributions of $\C_1$ with respect to contact
transformations.

\subsection{Bundles of Monge--Amp\`ere equations}
From now on we put $M = J^1\tau$.
\subsubsection{Bundles of hyperbolic Monge--Amp\`ere equations}
Let $\E$ be a Monge--Amp\`ere equation \eqref{MAeq}. It is
identified with the section
$$
  S_{\E} : x\mapsto\bigl[N(x):A(x):B(x):C(x):D(x)\bigr]
$$
of the trivial bundle
$$
     \rho:\RP^4\times M\longrightarrow M\,,\quad
     \bigl([p^0:p^1:p^2:p^3:p^4],\,x\bigr)\mapsto x\,,
$$
where $\RP^4$ is the 4-dimensional projective space. Obviously,
this identification is a bijection of the set of all
Monge--Amp\`ere equations onto the set of all sections of $\rho$.

Consider the open subset $E$ of the total space of $\rho$ defined
by the condition \eqref {Hprbl}, i.e.,
$$
  (p^2)^2-4p^1p^3+4p^4p^0 > 0\,.
$$
Clearly, the section $S_{\E}$ corresponding to a hyperbolic Monge--Amp\`ere
equation $\E$ takes values in $E$. Thus we can define the
bundle of hyperbolic Monge--Amp\`ere equations by the formula
\begin{equation} \label{pi}
  \pi=\rho\bigl|_E : E\longrightarrow M\,,\quad
     \bigl([p^0:p^1:p^2:p^3:p^4],\,x\bigr)\mapsto x\,.
\end{equation}

We use local coordinates $x,y,z,p,q$,
$u^1,\,\ldots,\,u^4$ in the total space $E$ of $\pi$,
where $x,y,z,p,q$ are the standard
coordinates on $M$, while the coordinates $u^1,\,\ldots,\,u^4$ on the
fibres of $\pi$ are defined as follows. Consider the
affine hyperplane in $\R^5$ defined by the equation $p^0=1$. It
generates the local chart in $E$
$$
     [1:p^1:p^2:p^3:p^4]\mapsto (p^1,p^2,p^3,p^4)\,.
$$
Following formulas \eqref{Dstr}, we introduce the local
coordinates $u^1,\,\ldots,\,u^4$ along the fibres of $\pi$ by
\begin{equation}\label{Coordnts0Tr}
     u^1=-p^3\,,\quad u^2=\frac{p^2-\sqrt\Delta}{2}\,,\quad
     u^3=\frac{p^2+\sqrt\Delta}{2}\,,\quad u^4=-p^1\,,
\end{equation}
where $\Delta=(p^2)^2-4p^1p^3+4p^4$.

These coordinates extend to the standard coordinates
$x,y,z,p,q,u^i,u^i_x,\allowbreak u^i_y,\allowbreak u^i_z,\allowbreak u^i_p,u^i_q$, \dots, $u^i_\sigma$, \dots, on $J^k\pi$, used in this paper
until we replace them with a more convenient set in Sect.~\ref{convenient}.

\subsubsection{The lifting of contact transformations}
Let $\varphi$ be a contact transformation defined in $M$. Then
$\varphi$ transforms any Monge--Amp\`ere equation $\E$ to another
Monge--Amp\`ere equation $\tilde\E$. In other words, $\varphi$
induces a transformation of the corresponding sections
$S_{\E}\mapsto S_{\tilde\E}$ and, consequently, a diffeomorphism
$\varphi^{(0)}$ of the total space of $\pi$ such that the diagram
$$
  \begin{CD}
    E @>\varphi^{(0)}>>  E\\
    @V\pi VV      @VV\pi V\\
    M             @>>\varphi>       M
  \end{CD}
$$
is commutative (in the domain of $\varphi^{(0)}$).
The
diffeomorphism $\varphi^{(0)}$ is called the {\it lifting of
$\varphi$ to the bundle $\pi$}.

The diffeomorphism $\varphi^{(0)}$, in its turn,
can be lifted to a diffeomorphism
$\varphi^{(k)}$ of $J^k\pi$ by the formula
$$
  \varphi^{(k)}(\,[S]^k_x\,)=\bigl[\,\varphi^{(0)}\circ
  S\circ\varphi^{-1}\,\bigr]^k_{\varphi(x)}\,.
$$
Obviously, for any $l>m$, the diagram
$$
  \begin{CD}
    J^l\pi         @>\varphi^{(l)}>> J^l\pi\\
    @V\pi_{l,m}VV             @VV\pi_{l,m} V\\
    J^m\pi         @>>\varphi^{(m)}> J^m\pi
  \end{CD}
$$
is commutative (in the domains of $\varphi^{(l)}$). The
diffeomorphism $\varphi^{(k)}$ is called the {\it lifting of
$\varphi$ to the jet bundle $J^k\pi$}.

\subsubsection{The lifting of contact vector fields}
\label{lifting}
Let $Z$ be a contact vector field in $M$ and let $\varphi_t$ be
its flow. Then $\varphi_t^{(k)}$ defines a vector field $Z^{(k)}$
in $J^k\pi$. This field is called the {\it lifting of $Z$ to
$J^k\pi$}. Obviously,
$$
  (\,\pi_{l,m}\,)_*\bigl(\,Z^{(l)}\,\bigr)=Z^{(m)}\,,\;\;
  \infty\geq l>m\geq -1\,,
$$
where $Z^{(-1)}=Z$.

It is not difficult to see that the map
$$
  Z\longmapsto Z^{(k)}
$$
is a homomorphism of the Lie algebra of all contact vector fields
onto the Lie algebra generated by all vector fields of the form
$Z^{(k)}$.

The local expression of $Z^{(k)}$ can be found as follows. First,
change the notation by putting
 $x^1=x$, $x^2=y$, $x^3=z$, $x^4=p$, $x^5=q$.
Recall that the operator $D_j$ of total derivative with respect to
$x^j$ in $J_{\infty}$ is given by the formula
$$
  D_j=\frac{\partial}{\partial x^j}
      +\sum_{|\sigma|\geq 0}\sum_{i=1}^{4}u^i_{\sigma j}
      \frac{\partial}{\partial u^i_{\sigma}}\,,\;j=1,2,\ldots,5\,,
$$
 The operator of evolution differentiation corresponding to a
generating function $\psi(Z)=(\psi^1(Z),\ldots,\psi^4(Z))^t$ is
defined by the formula
$$
 \re_{\psi(Z)}=\sum_{|\sigma|\geq 0}\sum_{i=1}^{4}
 D_{\sigma}\bigl(\,\psi^i(Z)\,\bigr)
 \frac{\partial}{\partial u^i_{\sigma}}\,,
$$
where $\sigma=\{j_1\ldots j_r\}\,,\;D_{\sigma}=D_{j_1}\circ\ldots
\circ D_{j_r}$ and $\psi(Z)$ is defined as follows.

Let $S$ be a section of $\pi$ defined in the domain of $Z$,
$\theta_1=[S]^1_x$, and $x=\pi_1(\theta_1)$; then
$$
  \psi(Z)(\theta_1)
  =\frac{d}{dt}(\,\varphi_t^{(0)}\circ S\circ
  \varphi_t^{-1}\,)\Bigr|_{t=0}(x)\,.
$$

If
$$
  Z=\sum_{i=1}^5 Z^i\frac{\partial}{\partial x^i}\,,
$$
then the lifting $Z^{(\infty)}$ is defined by the formula (see
\cite{KLV,KV})
\begin{equation}\label{Lift}
  Z^{(\infty)}=\sum_{j=1}^5Z^jD_j+\re_{\psi(Z)}\,.
\end{equation}
It follows from this formula that
\begin{equation}\label{kLift}
  Z^{(k)}=\sum_{j=1}^5Z^jD^k_j+\re^k_{\psi(Z)}\,,
\end{equation}
where
$$
  D^k_j=\frac{\partial}{\partial x^j}
  +\sum_{0\leq|\sigma|\leq k}\sum_{i=1}^{4}u^i_{\sigma j}
  \frac{\partial}{\partial u^i_{\sigma}}\,,\quad
  \re^k_{\psi(Z)}=\sum_{0\leq|\sigma|\leq k}\sum_{i=1}^{4}
  D_{\sigma}\bigl(\,\psi^i(Z)\,\bigr)
  \frac{\partial}{\partial u^i_{\sigma}}\,.
$$
Let $f$ be the generating function of the contact vector field $Z$
(see formula \eqref{CntVctFld}) and
$\theta_1=(\,x,y,z,p,q,u^i,u^i_x,u^i_y,u^i_z,u^i_p,u^i_q\,)$. Then
the vector $\psi(Z_f)(\theta_1)$ is $(\psi^1,\dots,\psi^4)$ with
\begin{equation}\label{DfmVlc}
\begin{array}{rcl}
\psi^1 &=& -u^1_zf
-u^1_p f_{ x}
-u^1_q f_{ y}
+(-p u^1_p-q u^1_q+u^1)f_{ z}\\&&\null
+(u^1_x+pu^1_z)f_{ p}
+(u^1_y+q u^1_z)f_{ q}
+f_{ x x}
+2pf_{ x z}
+p^2f_{ z z}\\&&\null
+2u^1f_{ x p}
+(u^2 + u^3)f_{ x q}
+2pu^1f_{ z p}
+p(u^2 + u^3)f_{ z q}\\&&\null
+(u^1)^2f_{ p p}
+(u^2+u^3)u^1f_{ p q}
+u^2u^3f_{ q q}
,
\\[6pt]
\psi^2 &=& -u^2_zf
-u^2_pf_{ x}
-u^2_qf_{ y}
+(-pu^2_p-qu^2_q+u^2)f_{ z}\\&&\null
+(u^2_x+pu^2_z)f_{ p}
+(u^2_y+qu^2_z)f_{ q}
+f_{ x y}
+qf_{ x z}
+pf_{ y z}
+pqf_{ z z}\\&&\null
+u^2f_{ x p}
+u^4f_{ x q}
+u^1f_{ y p}
+u^2f_{ y q}
+(q u^1+p u^2)f_{ z p}\\&&\null
+(qu^2+pu^4)f_{ z q}
+u^1u^2f_{ p p}
+(u^1u^4+(u^2)^2)f_{ p q}
+u^2u^4f_{ q q},
\\[6pt]
\psi^3 &=& -u^3_zf
-u^3_pf_{ x}
-u^3_qf_{ y}
+(-pu^3_p-qu^3_q+u^3)f_{ z}\\&&\null
+(u^3_x+pu^3_z)f_{ p}
+(u^3_y+qu^3_z)f_{ q}
+f_{ x y}
+qf_{ x z}
+pf_{ y z}
+pqf_{ z z}\\&&\null
+u^3f_{ x p}
+u^4f_{ x q}
+u^1f_{ y p}
+u^3f_{ y q}
+(qu^1+pu^3)f_{ z p}\\&&\null
+(qu^3+pu^4)f_{ z q}
+u^1u^3f_{ p p}
+(u^1u^4+(u^3)^2)f_{ p q}
+u^3u^4f_{ q q}
,
\\[6pt]
\psi^4 &=& -u^4_zf
-u^4_pf_{ x}
-u^4_qf_{ y}
+(-pu^4_p-qu^4_q+u^4)f_{ z}\\&&\null
+(u^4_x+pu^4_z)f_{ p}
+(u^4_y+qu^4_z)f_{ q}
+f_{ y y}
+2qf_{ y z}
+q^2f_{ z z}\\&&\null
+(u^2+u^3)f_{ y p}
+2u^4f_{ y q}
+q(u^2+u^3)f_{ z p}
+2qu^4f_{ z q}\\&&\null
+u^2u^3f_{ p p}
+(u^2+u^3)u^4f_{ p q}
+(u^4)^2f_{ q q}.
\end{array}
\end{equation}

\subsection{Differential invariants}
By $\Gamma$ we denote the pseudogroup of all contact
transformations of $M$. Its action is lifted to $J^k\pi, \;k\geq
0$, as it was explained above.

A function (vector field, differential form, or any other natural
geometric object on $J^k\pi$) is a {\it $k$\,th-order differential
invariant} of $\Gamma$ if for any $\varphi\in\Gamma$ the lifted
transformation $\varphi^{(k)}$ preserves this object. In this work
these differential invariants are called also {\it differential
invariants (of order~$k$) of Monge--Amp\`ere equations} or simply
{\it differential invariants (of order~$k$)}.

Let $\E$ be a Monge--Amp\`ere equation, $S_{\E}$ the section of
$\pi$ identified with $\E$, and $I$ a differential invariant of
order $k$. Then the {\it the value of $I$ on $\E$} is defined as
$(j_k S_{\E})^*(I) $ and denoted by $I_{\E}$. If a contact
transformation $f$ transforms $\E$ to $\tilde\E$, then, obviously,
$f^{(k)}$ transforms $I_{\E}$ to $I_{\tilde\E}$, for any $k$th
order invariant $I$.

Differential invariants that are functions are also called {\it
scalar differential invariants}. By $A_k$ we denote the
$\R$-algebra of all scalar differential invariants of order $\le
k$. By identifying $A_k$ with $\pi_{l,k}^*(A_k)\subset A_l,
\;\forall k\leq l,$ one gets a sequence of inclusions
$$
  A_0\subset A_1\subset\ldots\subset A_k\subset A_{k+1} \subset\ldots
$$
The $\R$-algebra $A=\bigcup_{k=0}^{\infty} A_k$ is called the {\it algebra
of scalar differential invariants of Monge--Amp\`ere equations}.

\begin{remark} \rm\label{restriction}
It is worth noticing that a scalar differential invariant $I$ is
completely determined by its values $I_{\E}$ on concrete equations
$\E$. This observation will be used below.
\end{remark}

Let $Z$ be a contact vector field in $M$ and $I$ a differential
invariant of order $k$. Then $L_{Z^{(k)}}(I)=0$, where $L$ stands
for the Lie derivative. This means, in particular, that $k$th
order scalar invariants are first integrals of all contact vector
fields lifted to $J^k\pi$. Obviously, a scalar differential
invariant of order $k$ is constant on any orbit  of the action of
$\Gamma$ on $J^k\pi$. Such an orbit consists, generally, of two
components, since contact transformations need not be orientation
preserving (e.g., the famous Legendre transformation $x' = p$, $y'
= q$, $z' = x p + y q - z$, $p' = x$, $q' = y$ is not). In other
words, the above-mentioned first integrals of $Z^{(k)}$ are,
generally, invariant  only with respect to the unit component of
$\Gamma$ and will be called {\it almost invariant}. Anyway,
generic orbits of contact transformations and of contact vector
fields have the same dimension:
\begin{proposition}
  \begin{enumerate}
    \item $J^k\pi$ is an orbit of the action of $\Gamma$ iff
          $k=0,1$,
    \item Codimension of a generic orbit of $J^2\pi$ is equal to
          $2$.
    \item Codimension of a generic orbit of $J^3\pi$ is equal to
          $29$.
   \end{enumerate}
\end{proposition}
\begin{proof} Let $\theta_k$ be a generic point of $J^k\pi$ and
$\Orb_{\theta_k}$ the orbit of the action of $\Gamma$ on $J^k\pi$
passing through $\theta_k$. Then $\codim\Orb_{\theta_k}=\dim
J^k\pi- \dim\Orb_{\theta_k}$. The dimension of $\Orb_{\theta_k}$
is the dimension of the subspace spanned by all vectors
$X^{(k)}(\theta_k)$ which can be calculated with the help of
computer algebra using formulas \eqref{kLift} and \eqref{DfmVlc}.
\end{proof}
An immediate consequence of the above proposition is
\begin{corollary}\label{NmIndScInv}
  \begin{enumerate}
    \item The algebra of scalar differential invariants $A_2$ is
          generated by $2$ functionally independent invarints.
    \item The algebra of scalar differential invariants $A_3$ is
          generated by $29$ functionally independent invarints.
   \end{enumerate}
\end{corollary}

Differential invariants constructed below come mainly form natural
geometric constructions without saying that these are invariant
with respect to the full pseudo-group~$\Gamma$. Although not
impossible, it is quite challenging task to obtain first integrals
of $Z^{(k)}$ analytically even for small $k$.

\section{Differential invariants on $J^2\pi$}

The next step to be done is explicit construction of differential
invariants that generate $A_2$ as a $C^{\infty}$-closed
algebra.
\subsection{Base projectors}
Let $\D$ be a distribution on $M$. Denote by $\D^{(1)}$ the
distribution generated by all vector fields $X$ and $[X, Y],
\;\forall \;X,Y \in \D$. Setting $\D^{(0)}=\D$, we define
$\D^{(r+1)}$, $r=0,1,\ldots$, inductively by the formula
$\D^{(r+1)}=(\D^{(r)})^{(1)}$.

\begin{lemma}\label{[]d}
For a hyperbolic Monge--Amp\`ere equation $\E$
$$
\dim(\D^1_\E)^{(1)} = \dim(\D^2_\E)^{(1)} = 3.
$$
\end{lemma}
\begin{proof}
Let $\omega\in \Lambda^1(M)$ and $X,Y\in D(M)$ be such that
$\omega(X)=\omega(Y)=0$. Then, by applying formula
$d\omega(X,Y)=L_X(Y\kch\,\omega)-L_Y(X\kch\,\omega)-[X,Y]\kch\,\omega$,
one easily finds that
$$
\omega([X,Y])=-d\omega(X,Y).
$$
If now $\omega=U_1$ and vector fields $X,Y\in\D_\E^i, \;i=1,2$,
are independent, then $dU_1(X,Y)\neq 0$ due to hyperbolicity of
$\E$. So, the above formula shows that $U_1([X,Y])\neq 0$, i.e.,
that $[X,Y]$ does not belong to the Cartan distribution on $M$.
So, $[X,Y]$ is independent on $X$ and $Y$.
\end{proof}
Restricting ourselves to the generic case only, we assume from now
on that
\begin{equation}
\label{generic}
\dim(\D^1_\E)^{(2)} = \dim(\D^2_\E)^{(2)} = 5\,.
\end{equation}

Suppose that vector fields $X_1$, $X_2$ generate the distribution
$\D^1_\E$ and vector fields $X_3$, $X_4$ generate the
distribution $\D^2_\E$. The 3-dimensional generic distributions
$\langle X_1, X_2, [X_1,X_2]\rangle$ and $\langle X_3, X_4, [X_3,
X_4]\rangle$ intersect along a one-dimensional subdistribution
$\D^3_\E=\langle X_1, X_2, [X_1, X_2]\rangle
    \cap\langle X_3, X_4, [X_3, X_4]\rangle$.
Hence, equation $\E$ generates a direct sum decomposition
\begin{equation}\label{DSD}
     T(M)=\D^1_\E\oplus\D^2_\E\oplus\D^3_\E.
\end{equation}
This decomposition generates six projections
$$
    \begin{aligned}
   \Prj_i &: T(M) \to \D^i_\E\,,\quad i=1,2,3\,,\\
   \Prj^{(1)}_j &: T(M) \to \D^i_\E \oplus \D^3_\E\,,\quad j=1,2\,,\\
   \Prj_\C &: T(M) \to \C = \D^1_\E\oplus\D^2_\E\,.\\
 \end{aligned}
$$

These projections may be viewed as vector-valued 1-forms. Namely,
let $X_5$ be a vector field generating $\D^3_\E$. Consider the
co-frame $\{\omega^1,\ldots,\omega^5\}$ on $M$ dual to the frame
$\{X_1,\ldots,X_5\}$, i.e., $\omega_i(X_j)=\delta_{ij}$. Then
\begin{equation}\label{Prjctrs}
    \begin{aligned}
       \Prj_1& = \omega^1\otimes X_1 + \omega^2\otimes X_2\,,\\
       \Prj_2&= \omega^3\otimes X_3 + \omega^4\otimes X_4\,,\\
       \Prj_3& = \omega^5\otimes X_5\,,\\
       \Prj^{(1)}_j& = \Prj_j + \Prj_3\,, \quad j = 1,2\,,\\
       \Prj_\C &= \Prj_1 + \Prj_2.
 \end{aligned}
\end{equation}

These vector-valued differential 1-forms are, obviously,
differential invariants of $\E$ with respect to contact
transformations. Moreover, according to proposition
\ref{reconstr}, the original equation $\E$ is completely
determined by each of the projectors $\Prj_1$, $\Prj_2$.

\subsection{Coordinate-wise description of base projectors}
In order to find local expressions for the above projectors,
consider vector fields $X_1,\ldots,X_4$ given by \eqref{Dstr} and
use the notation \eqref{Coordnts0Tr}, i.e.,
\begin{equation}\label{Y1-Y4}
     \begin{aligned}
      X_1&=\frac{\partial}{\partial x}
         +p\frac{\partial}{\partial z}
         +u^1\frac{\partial}{\partial p}
         +u^2\frac{\partial}{\partial q}\,,\;
       X_2=\frac{\partial}{\partial y}
         +q\frac{\partial}{\partial z}
         +u^3\frac{\partial}{\partial p}
         +u^4\frac{\partial}{\partial q}\,,\\
       X_3&=\frac{\partial}{\partial x}
         +p\frac{\partial}{\partial z}
         +u^1\frac{\partial}{\partial p}
         +u^3\frac{\partial}{\partial q}\,,\;
       X_4=\frac{\partial}{\partial y}
         +q\frac{\partial}{\partial z}
         +u^2\frac{\partial}{\partial p}
         +u^4\frac{\partial}{\partial q}\,.
     \end{aligned}
\end{equation}

The remaining field $X_5$ is defined by the
relation
\begin{equation}\label{Y5}
     X_5 = \lambda^1X_1 + \lambda^2X_2 + \kappa[X_1, X_2]
         = \lambda^3X_3 + \lambda^4X_4 + \chi[X_3, X_4]\,.
\end{equation}
A simple computation shows that
$$
     \lambda^3 = \lambda^1\,,\quad
     \lambda^4 = \lambda^2\,,\quad
     \chi = -\kappa \neq 0\,,
$$
with
\begin{equation}\label{lambda}
\begin{aligned}
     \lambda^1 = \frac{1}{u^2-u^3}
     \Bigl(&(u^2+u^3)_y+q(u^2+u^3)_z+u^4(u^2+u^3)_q\\
           &\null -2(u^4_x+pu^4_z+u_1u^4_p)-(u^2+u^3)u^4_q+u^3u^2_p+u^2u^3_p\Bigr),
\\
     \lambda^2 = \frac{1}{u^2-u^3}
     \Bigl(&(u^2+u^3)_x+p(u^2+u^3)_z+u^1(u^2+u^3)_p\\
           &\null -2(u^1_y+qu^1_z+u_4u^1_q)-(u^2+u^3)u^1_p+u^2u^3_q+u^3u^2_q\Bigr)\,
\end{aligned}
\end{equation}
provided that  $X_5$ is normalized by the requirement $\kappa =
1$.

Brackets of vector fields $X_1,\ldots,X_5$ are described by means
of the coefficients $b^i_{jk}$:
$$
[X_j,X_k] = \sum_{i=1}^5 b^i_{jk}X_i\,.
$$
Obviously,
$b^i_{jk}=-b^i_{kj}$.

\subsection{Convenient coordinates on $J^k\pi$}
\label{convenient}
Vector fields $X_i$, $i = 1,\dots,5$ induce vector fields $\X_i$
on the bundle $J^{\infty}\pi$, uniquely defined by the condition
$j^k(S_\E)_*X_i = \X_i$ for all sections $S_\E$. Thus, $\X_1 = D_1
+ p D_3 + u^1 D_4 + u^2 D_5$, etc., where $D_i$ denote the total
derivatives, see Sect.~\ref{lifting}.

Differential invariants of hyperbolic Monge--Amp\`ere equations
constructed bellow are described in terms of the quantities
$\X_{i_1} \ldots \X_{i_h} b^k_{ij}$. So, we need to know all
algebraic relations connecting them, at least for $h = 0,1$. To
find these efficiently it is convenient to use a non-standard
local chart in $J^k\pi$.
\begin{lemma}
Functions
\begin{equation}\label{newcoord}
  \tilde u^j_{i_1\ldots i_h} = \X_{i_1} \ldots \X_{i_h} u^j, \quad
  i_1 \le \ldots \le i_h, \;h \le k.
\end{equation}
together with functions $x^i, u^j$ constitute a local chart on
$J^k\pi$. Moreover, the standard jet coordinates on $J^k\pi$ are
rational functions of these coordinates.
\end{lemma}
\begin{proof}
For $k = 2$ the assertion is verified directly. For $k
> 2$ one can express the standard jet coordinates 
$u^j_{i_1\ldots i_k} = D_{i_1\ldots i_k} u^j$ 
in terms of coordinates \eqref{newcoord} by making use
of the following obvious facts. First, fields $D_i$ are linear
combinations of fields $\X_i$ with coefficients in $C^\infty
(J^2\pi)$. Second, the coefficients $b^j_{i_1 i_2}$ are functions
on $J^2\pi$. Third, $\X_{i_2}\X_{i_1} f = -b^j_{i_1 i_2} \X_j f +
\X_{i_1}\X_{i_2} f$ for every function $f\in C^\infty (J^k\pi),
\;k \ge 2$.
\end{proof}

A complete system of relations connecting functions $b^k_{ij}$ can
be found by routine computations and taking into consideration
geometric properties of fields $X_1,\dots,X_5$. For instance,
$b^3_{12}=b^4_{12}=0$, since $[X_1,X_2]$ belongs to the
distribution $(\D^1_\E)^{(1)}$ generated by $X_1,X_2$ and $X_5$,
etc. The final result is as follows:
\begin{equation}\label{Brckts1}
\def\arraystretch{1.7}
\def\b#1#2#3{b^#3_{#1#2}}
\begin{array}{l}
\begin{array}{lllll}
\b341 = 0, &
\b342 = 0, \\
\b123 = 0, &
\b124 = 0, &
\b125 = 1, \\
\b133 = -\b131, &
\b134 = -\b132, &
\b135 = 0, \\
\b233 = -\b231, &
\b234 = -\b232, &
\b235 = 0, \\
\b143 = -\b141, &
\b144 = -\b142, &
\b145 = 0, \\
\b243 = -\b241, &
\b244 = -\b242, &
\b245 = 0, \\
\b343 = -\b121, &
\b344 = -\b122, &
\b345 = -1,
\end{array}
\\
\begin{array}{ll}
\b155 = -\b142-\b131, &
\b255 = -\b242-\b231, \\
\b355 = -\b131-\b232, &
\b455 = -\b242-\b141, \\
\b454 = -\b353+\b252+\b151, \\
\end{array}
\end{array}
\end{equation}


Henceforth we shall simplify the notation by using $X_i$ for
$\X_i$.


\subsection{Curvatures}

Using formulas \eqref{Omega}, \eqref{Crv} and
 the direct sum decomposition \eqref{DSD}, it is easy to compute
the curvature forms of projectors $\Prj_1$, $\Prj_2$,
$\Prj^{(1)}_1$, $\Prj^{(1)}_2$, $\Prj_\C$, which are
\begin{equation}\label{Crvtrs}
\begin{aligned}
  \Crv_1 &= \omega^1\wedge\omega^2\otimes X_5\,,\\ 
  \Crv_2 &= -\omega^3\wedge\omega^4\otimes X_5\,,\\ 
  \Crv_1^1&=-(b^3_{15}\omega^1+b^3_{25}\omega^2)\wedge\omega^5
  \otimes X_3
            -(b^4_{15}\omega^1+b^4_{25}\omega^2)\wedge\omega^5
  \otimes X_4\,,\\
  \Crv_2^1&=-(b^1_{35}\omega^3+b^1_{45}\omega^4)\wedge\omega^5
  \otimes X_1
            -(b^2_{35}\omega^3+b^2_{45}\omega^4)\wedge\omega^5
  \otimes X_2\,,\\
    \Crv&=\Crv_1+\Crv_2\,,
\end{aligned}
\end{equation}
respectively. It is clear that these curvature forms are
differential invariants of $\E$.

Fr\"olicher--Nijenhuis brackets of base projectors give new
invariant vector-valued forms. These, however, turn out to be
linear combinations of curvature forms. More exactly, a direct
computation, which is omitted, shows that
\begin{align*}
    [\![\Prj_1, \Prj_2]\!]&=\tfrac{1}{2}(-[\![\Prj_1, \Prj_1]\!]
    -[\![\Prj_2, \Prj_2]\!]+[\![\Prj_3, \Prj_3]\!])\,,\\
    [\![\Prj_1, \Prj_3]\!]&=\tfrac{1}{2}(-[\![\Prj_1, \Prj_1]\!]
    +[\![\Prj_2, \Prj_2]\!]-[\![\Prj_3, \Prj_3]\!])\,,\\
    [\![\Prj_2, \Prj_3]\!]&=\tfrac{1}{2}([\![\Prj_1, \Prj_1]\!]
    -[\![\Prj_2, \Prj_2]\!]-[\![\Prj_3, \Prj_3]\!])
\end{align*}
and
\begin{align*}
    [\![\Prj_1, \Prj_1]\!]&=-2(\Crv_2^1+\Crv_1),\quad
    [\![\Prj_2, \Prj_2]\!]=-2(\Crv_1^1+\Crv_2)\,,\\
    [\![\Prj_3, \Prj_3]\!]&=-2(\Crv_1+\Crv_2).
\end{align*}


\subsection{Scalar invariants on $J^2\pi$}
The following three invariant 5-forms with values in $\D^3_{\E} =
\langle X_5\rangle$:
\begin{equation}\label{V512a}
  \begin{aligned}
    &\tfrac{1}{2}\bigl(\Crv_2^1\kch\Crv_1\bigr)
    \kch\bigl(\Crv_2^1\kch\Crv_1\bigr)
    =\Lambda_1\,
            \omega^1\wedge\ldots\wedge\omega^5\otimes X_5\,,\\
    &\tfrac{1}{2}\bigl(\Crv_1^1\kch\Crv_2\bigr)
    \kch\bigl(\Crv_1^1\kch\Crv_2\bigr)
    =\Lambda_2\,
            \omega^1\wedge\ldots\wedge\omega^5\otimes X_5\,,\\
    &\bigl(\Crv_2^1\kch\Crv_1\bigr)
    \kch\bigl(\Crv_1^1\kch\Crv_2\bigr) =\Lambda_{12}\,
    \omega^1\wedge\ldots\wedge\omega^5\otimes X_5\,,
     \end{aligned}
\end{equation}
with
\begin{equation}
\label{Lambda}
\begin{gathered}
\Lambda_1 = b^2_{35}b^1_{45} - b^1_{35}b^2_{45},\qquad
\Lambda_2 = b_{15}^4b_{25}^3 - b_{15}^3b_{25}^4,\\
\Lambda_{12} = b_{15}^3b_{35}^1 + b_{15}^4b_{45}^1
    + b_{25}^3b_{35}^2 + b_{25}^4b_{45}^2\,.
\end{gathered}
\end{equation}
are proportional. Therefore, the corresponding proportionality
factors are scalar differential invariants. In particular, such
are
\begin{equation}\label{ScInv12}
   \begin{aligned}
    I^1&=\Lambda_{12}/\Lambda_1\,,\\
    I^2&=\Lambda_{12}/\Lambda_2\;.
     \end{aligned}
\end{equation}
Below it will be shown that $\Lambda_1, \Lambda_2$ are nowhere
zero.

\begin{theorem}
  The algebra of scalar differential invariants on $J^2\pi$
  is generated by the invariants $I^1$ and $I^2$.
\end{theorem}
\begin{proof} In view of Corollary \ref{NmIndScInv}, it is sufficient to show
that $I^1$ and $I^2$ are functionally independent (on $J^2\pi$).
But this is straightforward from \eqref{Brckts1}.
\end{proof}

Coefficients $\Lambda_{\sigma}, \;\sigma =1,2,12$, introduced in
\eqref{Lambda} have a geometrical meaning explained below.
Fix a generator $W = f X_5$ in $\D^3_{\E}$ and consider maps
$$
\square_1^W : \D^2 \to \D^1, \qquad \square_2^W : \D^1 \to \D^2,
$$
defined by formulas
$$
\square_1^W(Z_2) = \Prj_1([Z_2,W]), \qquad \square^W_2(Z_1) =
\Prj_2([Z_1,W]),
$$
with $Z_1 \in \D^1$, $Z_2 \in \D^2$. Since
$\Prj_1(\D^2_{\E})=\Prj_2(\D^1_{\E})=0$ both $\square_1^W$ and
$\square_1^W$ are $C^\infty(M)$-linear. This is seen as well from
their local expressions
\begin{eqnarray*}
\square_1^W = f b^i_{j5}\,\omega^j \otimes X_i, & i = 1,2, & j =
3,4,
\\
\square_2^W = f b^i_{j5}\,\omega^j \otimes X_i, & i = 3,4, & j =
1,2.
\end{eqnarray*}
Consider also 2-forms $\rho_i^W : \D^i \times \D^i \to \R$, $i =
1,2$, defined by
\begin{equation}
\label{rho} \rho_i^W(U_i,V_i) W = \Crv_i(U_i,V_i), \quad U_i,V_i
\in \D^i_{\E}.
\end{equation}
Then, obviously, $\rho_1^W = (1/f) \omega^1 \wedge \omega^2$,
$\rho_2^W = -(1/f) \omega^3 \wedge \omega^4$, so that both
are volume forms of $\D^1$ and $D^2$, respectively. Moreover, we
have
\begin{equation}
\label{chi}
\begin{gathered}
(\square_1^W)^* (\rho_1^W) = 
f^2\Lambda_2 \rho_2^W,
\\
(\square_2^W)^* (\rho_2^W) = 
f^2\Lambda_1 \rho_1^W,
\\
\tr(\square_1^W \circ \square_2^W) = \tr(\square^W_2 \circ
\square^W_1)
 = f^2 \Lambda_{12}.
\end{gathered}
\end{equation}

\begin{proposition}
\label{nonzero} If $\E$ is generic, then functions $\Lambda_1$,
$\Lambda_2$ are nowhere zero.
\end{proposition}

\begin{proof}
By genericity condition \eqref{generic}, $\square_1^W$ and
$\square^W_2$ are surjective, hence $\Lambda_1,\Lambda_2$ are
nonzero.
\end{proof}

\subsubsection{}
\label{subtype}

Now consider operators $\nabla_1^W = \square^W_1 \circ
\square^W_2$ and $\nabla_2^W = \square^W_2 \circ \square^W_1$
acting on $\D^1$ and $\D^2$, respectively. It follows from
\eqref{chi} that
\begin{equation}\label{char}
\lambda^2 - f^2 \Lambda_{12} \lambda + f^4 \Lambda_1 \Lambda_2
\end{equation}
is the characteristic polynomial for each of them. Another
peculiarity of the situation is that $\square_1^W$ send
eigenvectors of $\nabla_2^W$ to that of $\nabla_1^W$ and similarly
for $\square_1^W$.

The discriminant of polynomial (\ref{char}) is
$$
f^4 \Lambda_1 \Lambda_2 (I^1 I^2 - 4).
$$
Its sign coincides, obviously, with the sign of
$$
I^1 I^2(I^1 I^2 - 4)\;.
$$
This proves that generic hyperbolic Monge--Amp\`ere equations are
subdivided into three subclasses as follows:
  \begin{enumerate}
  \item subclass ``h": the operator $\nabla_i^W$ has two different real
      eigenfunctions \\ $\Leftrightarrow$ \quad $I^1 I^2(I^1 I^2 - 4)>0$,
      \item subclass ``p": the operator $\nabla_i$ has a unique real
    eigenfunction\\
      $\Leftrightarrow$ \quad $I^1 I^2(I^1 I^2 - 4)=0$,
    \item subclass ``e": the operator $\nabla_i$ has no real
      eigenfunctions\\ $\Leftrightarrow \quad I^1 I^2(I^1 I^2 - 4)<0$.
  \end{enumerate}

\subsubsection{Some almost invariants}
\label{ngi-1} The previous considerations lead to an almost
invariant choice of generator $W = f X_5$ in $\D^3_{\E}$. Namely,
define functions $\Lambda_i^W, \;i=1,2$, by relations
$$
(\square_1^W)^* (\rho_1^W) = \Lambda_2^W \rho_2^W, \quad
(\square_2^W)^* (\rho_2^W) = \Lambda_1^W \rho_1^W\;.
$$
Obviously, $\Lambda_i^W=f^2\Lambda_i$. This shows that,  up to
sign, vector fields
$$
W_i=\frac{1}{\sqrt{|\Lambda_i^W|}}W, \quad i=1,2,
$$
do not depend on the choice of $W$. In particular,
$\Lambda_i^{X_5}=\Lambda_i$, so that
$$
W_i=\frac{1}{\sqrt{|\Lambda_i|}}X_5, \quad i=1,2.
$$
By duality, 1-forms
$$
\vartheta_i=\sqrt{|\Lambda_i|}\omega_5, \;i=1,2,
$$
are almost invariant as well.

It is not difficult to construct further almost invariant forms.
For instance, the forms
$$
  \vartheta_{ij}=\Crv_i\kch\vartheta_j,\;i=1,2,
$$
are manifestly almost invariant and have the following local
expressions:
$$
\vartheta_{1j}=\sqrt{|\Lambda_j|}\omega^1\wedge\omega^2 \quad
\vartheta_{2j}=\sqrt{|\Lambda_j|}\omega^3\wedge\omega^4\;. $$

The products
\begin{equation}
\label{rho1,rho2} \rho_j = (\mathop{-\mathrm{sign}}\Lambda_j)
\vartheta_{1j} \wedge \vartheta_{2j}
 = \Lambda_j \,\omega^1 \wedge \omega^2 \wedge \omega^3 \wedge \omega^4,  \qquad j = 1,2,
\end{equation}
which are volume forms on the Cartan distribution $\D^1_{\E},
\oplus \D^2_{\E}$, are, obviously, fully invariant. This is a very
simple example on how an invariant can be constructed from almost
invariants. Forms $\rho_j$ can be described in a manifestly
invariant way as follows:
$$
\rho_1=\frac{1}{2}\bigl\langle\,(\Crv_2^1\kch\Crv_1)\kch
(\Crv_2^1\kch\Crv_1)\,\bigr\rangle, \quad
\rho_2=\frac{1}{2}\bigl\langle\,(\Crv_1^1\kch\Crv_2)\kch
(\Crv_1^1\kch\Crv_2)\,\bigr\rangle
$$
where $\langle\,\cdot,\cdot\,\rangle$ stands for convolution. Note that
the form
\begin{equation}
\label{rho12} \rho^{12} =I^j\rho_j=\Lambda_{12} \,\omega^1 \wedge
\omega^2 \wedge \omega^3 \wedge \omega^4
\end{equation}
is invariant too.

Similarly, one can construct many other invariant forms. Some of
them are :
\begin{equation}
\begin{aligned}
\bigl\langle\,\Crv_2^1\kch\Crv_1\,\bigr\rangle=&-(b^1_{35}\omega^3+b^1_{45}\omega^4)\wedge\omega^2
     +(b^2_{35}\omega^3+b^2_{45}\omega^4)\wedge\omega^1\\
\bigl\langle\,\Crv_1^1\kch\Crv_2\,\bigr\rangle=&(b^3_{15}\omega^1+b^3_{25}\omega^2)\wedge\omega^4
  -(b^4_{15}\omega^1+b^4_{25}\omega^2)\wedge\omega^3\,\\
\Crv_1^1\kch&\bigl\langle\,\Crv_2\kch\Crv_1^1\,\bigr\rangle=2\Lambda_2\,
 \omega^1\wedge\omega^2\wedge\omega^5\\
\Crv_2^1\kch&\bigl\langle\,\Crv_2^1\kch\Crv_1\,\bigr\rangle=2\Lambda_{1}\,
 \omega^3\wedge\omega^4\wedge\omega^5\\
\end{aligned}
\end{equation}

Now it is easy to construct almost invariant volume forms :
\begin{equation}
\label{xi} \vartheta_j\wedge\rho_j = |\Lambda_j|^{3/2}\, \omega^1
\wedge \omega^2 \wedge \omega^3 \wedge \omega^4 \wedge \omega^5,
\quad j = 1,2.
\end{equation}

\section{Differential invariants on $J^3\pi$}
%

Since $\omega^k(X_l)=const$, namely, $\delta_{kl}$, we have
$$d\omega_k(X_i,X_j)=-\omega^k ([X_i,X_j])\;.$$
(see the proof of lemma \ref{[]d}). This implies the useful
formula
\begin{equation}\label{d-omega}
d\omega^k=-\sum_{i<j}b^k_{ij}\,\omega^i\wedge\omega^j\,.
\end{equation}

\subsection{The complete parallelism}
\label{parallelism}
First, note that invariant differential 1-forms $dI^1$ and $dI^2$
live on $J^3\pi$. This leads us immediately to another set of
invariant differential 1-forms on $J^3\pi$:
\begin{equation}\label{CmplPrll}
\begin{aligned}
   \Omega^1&=\Prj_1\kch dI^1=X_1(I^1)\,\omega^1+X_2(I^1)\,\omega^2\,,\\
   \Omega^2&=\Prj_1\kch dI^2=X_1(I^2)\,\omega^1+X_2(I^2)\,\omega^2\,,\\
   \Omega^3&=\Prj_2\kch dI^1=X_3(I^1)\,\omega^3+X_4(I^1)\,\omega^4\,,\\
   \Omega^4&=\Prj_2\kch dI^2=X_3(I^2)\,\omega^3+X_4(I^2)\,\omega^4\,,\\
   \Omega^5_1&=\Prj_3\kch dI^1=X_5(I^1)\,\omega^5\,,\quad
   \Omega^5_2=\Prj_3\kch dI^2=X_5(I^2)\,\omega^5\,.
\end{aligned}
\end{equation}

Supposing that $\E$ is a generic equation, we henceforth assume that
\begin{equation}\label{Assmng1}
   X_5(I^1)\neq 0\,,\quad  X_5(I^2)\neq 0\,,
\end{equation}
and
\begin{equation}\label{Assmng2}
   \Delta_1=
      \begin{vmatrix}
          X_1(I^1) & X_2(I^1)\\
          X_1(I^2) & X_2(I^2)
     \end{vmatrix}\neq 0\,,\quad
   \Delta_2=
      \begin{vmatrix}
          X_3(I^1) & X_4(I^1)\\
          X_3(I^2) & X_4(I^2)
     \end{vmatrix}\neq 0\,.
\end{equation}
This means that two sets of forms $\{\Omega^1,\ldots,\Omega^4,
\Omega^5_1\}$ and $\{\Omega^1,\ldots,\Omega^4$, $\Omega^5_2\}$ are
invariant coframes on $M$ (we omit the subscript $\E$ according to
Remark \ref{restriction}). Each of these coframes determines an
invariant complete parallelism on $M$.

The frames $\{Y_1,\ldots$, $Y_4$, $Y_5^1\}$ and
$\{Y_1,\ldots,Y_4,Y_5^2\}$, dual to the above constructed
coframes, are, obviously, invariant. An explicit description of
them is :
\begin{equation}\label{CmplPrll1}
\begin{aligned}
   Y_1&=\frac{1}{\Delta_1}\bigl(X_2(I^2)X_1-X_1(I^2)X_2\bigr)\,,\\
   Y_2&=\frac{1}{\Delta_1}\bigl(-X_2(I^1)X_1+X_1(I^1)X_2\bigr)\,,\\
   Y_3&=\frac{1}{\Delta_2}\bigl(X_4(I^2)X_3-X_3(I^2)X_4\bigr)\,,\\
   Y_4&=\frac{1}{\Delta_2}\bigl(-X_4(I^1)X_3+X_3(I^1)X_4\bigr)\,,\\
   Y_5^1&=\frac{1}{X_5(I^1)}X_5\,,\quad
   Y_5^2=\frac{1}{X_5(I^2)}X_5\,.
\end{aligned}
\end{equation}

\subsection{More scalar invariants on $J^3\pi$}
%
Among numerous invariants constructed previously there are
functions, (vector-valued) differential forms,and vector fields.
Further invariants can by obtained just by applying various
operations of tensor algebra, Fr\"olicher--Nijenhuis brackets,
etc, to these objects. Moreover, components of an invariant object
with respect to an invariant basis are scalar differential
invariants as well as its proper differential invariants. These
simple general tricks are rather efficient and were already used
in constructing differential invariants on $J^2\pi$. As for
$J^3\pi$ we shall proceed along these lines as well.

The invariant 1-forms $\Omega^5_1$ and $\Omega^5_2$ are
proportional. So, the proportionality factor
\begin{equation}\label{I3}
   I^3=\frac{X_5(I^1)}{X_5(I^2)}
\end{equation}
is a scalar differential invariant on $J^3\pi$.

Consider now invariant 2-forms on $J^3\pi$:
\begin{equation}\label{I's}
\begin{aligned}
   \Crv_1\kch dI^1&=I^6\Omega^1\wedge\Omega^2\,,\\
   \Crv_1\kch dI^2&=I^7\Omega^1\wedge\Omega^2\,,\\
   \Crv_2\kch dI^1&=I^8\Omega^3\wedge\Omega^4\,,\\
   \Crv_2\kch dI^2&=I^9\Omega^3\wedge\Omega^4\,,\\
   \Crv_1^1\kch dI^1&=I^{10}\Omega^1\wedge\Omega^5_1+I^{11}\Omega^2\wedge\Omega^5_1\,,\\
   \Crv_1^1\kch dI^2&=I^{12}\Omega^1\wedge\Omega^5_1+I^{13}\Omega^2\wedge\Omega^5_1\,,\\
   \Crv_2^1\kch
dI^1&=I^{14}\Omega^3\wedge\Omega^5_1+I^{15}\Omega^4\wedge\Omega^5_1\,,\\
   \Crv_2^1\kch
dI^2&=I^{16}\Omega^3\wedge\Omega^5_1+I^{17}\Omega^4\wedge\Omega^5_1\,.
\end{aligned}
\end{equation}

Their components $I^6,\dots,I^{17}$ with respect to the base
$\Omega^1,\dots,\Omega^5$ are further scalar differential
invariants on $J^3\pi$. The simplest among them are
$I^6={\Delta_1}/{X_5(I^1)}$ and $I^8={\Delta_2}/{X_5(I^1)}$.

In the same manner one easily find  numerous  non-scalar
differential invariants on $J^3\pi$. For instance, such are
3-forms $[\![\Prj_i, \Crv_j]\!]$ or $[\![\Prj_i, \Crv_j^1]\!]$,
4-forms $[\![\Prj_i, (\Crv_j^1\kch \Crv_k^1)]\!]$, 5-forms
$[\![\Prj_i, \Crv_j^1]\!]\kch [\![\Prj_k, \Crv_l^1]\!]$, etc.

\subsection{Better manageable invariants}\label{ngi-2}
From the above said one can see that there are
sufficient resources for constructing differential invariants and
the main problem becomes to select functionally
independent ones in the simplest possible way. From technical point of
view this forces us to look for {\it manageable} invariants, for
instance, those that have local expression as simple as possible.
In the considered context a help comes from almost invariant
objects as it is illustrated below.


In view of \eqref{rho1,rho2},  \eqref{rho12} and \eqref{d-omega},
for $\sigma = 1,2,12$ we have the invariant 5-forms
\begin{equation}\label{Df-3form}
\begin{aligned}
d\rho_{\sigma} & = d(\Lambda_\sigma \,\omega^1 \wedge \omega^2
\wedge \omega^3 \wedge \omega^4)
\\
& = \bigl(X_5(\Lambda_\sigma) + \Lambda_\sigma B \bigr) \omega^1
\wedge \omega^2 \wedge \omega^3 \wedge \omega^4 \wedge \omega^5,
\end{aligned}
\end{equation}
where $B = b^1_{15} + b^2_{25} + b^3_{35} + b^4_{45} = 2(b^1_{15}
+ b^2_{25}) =  2(b^3_{35} + b^4_{45})$ according to
\eqref{Brckts1}.

By comparing these 5-forms with \eqref{xi} we obtain almost scalar
invariants
\begin{equation}
\label{Ijsigma} I^j_\sigma = \frac{X_5(\Lambda_\sigma) +
\Lambda_\sigma B} {|\Lambda_j|^{3/2}}, \qquad \sigma = 1,2,12, \ \
j = 1,2,
\end{equation}
on $J^3\pi$ which are better manageable in comparison to those
constructed in the previous subsection. The squares
$(I^j_\sigma)^2$ are, obviously, full scalar invariants. Apart
from the obvious relation $(I^1_\sigma/I^2_\sigma)^2 =
(I^1/I^2)^3$ they are functionally independent. Some of the
earlier constructed invariants can be expressed in terms of
$I^j_\sigma$'s, e.g.,
$$
\frac{X_5(I^1)}{X_5(I^2)} =
  \frac{(I^j_{12} - I^j_1 I^1)I^1}{(I^j_{12} - I^j_2 I^2)I^2}, \quad j =1,2.
$$

\section{The equivalence problem}
So far we obtained two independent second-order scalar invariants
$I^1,I^2$ \;(see~\eqref{ScInv12}) and a number of third-order
invariants. Put (see~\eqref{Ijsigma})
$$
I^3 = (I^1_1)^2, \qquad
I^4 = (I^1_2)^2, \qquad
I^5 = (I^1_{12})^2,
$$
The following statement can be checked by a direct computer-supported
calculation in coordinates \eqref{newcoord}:
\begin{theorem}\label{Indpndns}
   Invariants $I^1$, $I^2$, $I^3$, $I^4$, and $I^5$ are
   functionally independent on $J^3(\pi)$.
\end{theorem}

Of course, this choice of basic scalar invariants is not unique.
For instance, invariants $I^1, I^2, I^3, I^6, I^8$ (see
\eqref{I3}, \eqref{I's}) are functionally independent as well.
However, this and other reasonable choices are "less manageable"
with respect to those made in the above theorem. Unfortunately, this
fact is not clearly seen from the above exposition, since we were
forced to skip technical details of computations.

 According to ``the principle of $n$ invariants"
\cite{V1}, any quintuple of functionally independent scalar
invariants gives a solution of the equivalence problem for generic
hyperbolic Monge--Amp\`ere equations. Theorem \ref{Indpndns}
guarantees existence of a such one, namely, $I^1,\ldots,I^5$.

More exactly, let $\E$ be a generic hyperbolic Monge--Amp\`ere
equation considered as a section of the bundle $\pi$. Since
invariants $I^1,\ldots,I^5$ are functionally independent their
values $I^1_{\E},\ldots,I^5_{\E}$ on $\E$ form a (local) chart in
$M$. In terms of these coordinates, the 1-forms
$\Omega_1,\ldots,\Omega_5$, defining an absolute parallelism on
$M$, are described in terms of functions
$\Omega^i_j(I^1_{\E},\ldots,I^5_{\E})$ coming from the
decomposition
$$
   \Omega_i=\sum_{j=1}^5\Omega^i_j(I^1_{\E},\ldots,I^5_{\E})dI^j_{\E}\,,
   \quad i=1,\ldots,5\,.
$$
\begin{theorem}
   The (local) equivalence class of a generic equation $\E$ with respect to contact
   transformations is uniquely determined by the family of functions
   $\Omega^i_j(I^1_{\E},\ldots,I^5_{\E}), \;i=1,\ldots,5\,.$
\end{theorem}
\begin{proof} Let $\tilde\E$ be another generic Monge--Amp\`ere
equation such that there exists a contact transformation
transforming it to $\E$. Then, obviously, the functions
$\Omega^i_j(I^1_{\E},\ldots,I^5_{\E})$ and
$\tilde\Omega^i_j(I^1_{\tilde\E},\ldots,I^5_{\tilde\E})$ coincide
for all $i$ and $j$.

Let $\E,\tilde\E$ be Monge--Amp\`ere equations such that  for all
$i$ and $j$ the functions $\Omega^i_j(I^1_{\E},\ldots,I^5_{\E})$
and $\tilde\Omega^i_j(I^1_{\tilde\E},\ldots,I^5_{\tilde\E})$
coincide. Let $I_{\E}=(I^1_{\E},\ldots,I^5_{\E})$ and
$I_{\tilde\E}=(I^1_{\tilde\E},\ldots,I^5_{\tilde\E})$ be invariant
coordinate systems in $M$ for $\E$ and $\tilde\E$ respectively.
Then $I_{\tilde\E}^{-1} \circ I_{\E}$ is a locally defined
diffeomorphism $M \to M$. This diffeomorphism is a contact
transformation because it transforms
$\Omega_i=\sum_{j=1}^5\Omega^i_j(I^1_{\E},\ldots,I^5_{\E})dI^j_{\E}$
to
$\sum_{j=1}^5\tilde\Omega^i_j(I^1_{\tilde\E},\ldots,I^5_{\tilde\E})d
I^j_{\tilde\E}=\tilde\Omega^5, \;i=1,\ldots,5$, and, in
particular, the contact form $\Omega_5$ to the contact form
$\tilde{\Omega_5}$. By obvious reasons it also transforms the pair
of distributions $(\D^1_{\E}, \D^2_{\E})$ to the pair
$(\D^1_{\tilde\E}, \D^2_{\tilde\E})$ and hence $\E$ to
$\tilde{\E}$.
\end{proof}


\section{Examples}

Examples discussed in this section aim to illustrate the character
and complexity of problems related with actual computations and
use of differential invariants. Henceforth invariants $I^i$ are
denoted by $I_i$.

\begin{example} \rm
Consider the equation
$$
\textstyle\frac{1}{4} (z_{xx} z_{yy} - z_{xy}^2)
 + y^2 z_{xx} - 2 x y z_{xy} + x^2 z_{yy} + x^2 y^2 z^2 = 0.
$$
The first two invariants are $I_1 = z n_+/d$, $I_2 = z n_-/d$, where
\begin{eqnarray*}
n_{\pm}
&=&
2 (z + 3 y^4 \mp 2) x^2 z_x^2
 - (z + 12 x^2 y^2) x y z_x z_y
 + 2 (z + 3 x^4 \pm 2) y^2 z_y^2
\\&&\null
 + (z^2 + 8 x^2 y^2 z + 4 y^4 z \pm 4 z \pm 16 x^2 y^2 \pm 16 y^4 - 12) x z_x
\\&&\null
 + (z^2 + 4 x^4 z + 8 x^2 y^2 z \mp 4 z \mp 16 x^4 \mp 16 x^2 y^2 - 12) y z_y
\\&&\null
 + 2 z^3 + 36 x^4 y^4 z^3
 + 6 y^4 z^2 - 4 x^2 y^2 z^2 + 6 x^4 z^2
\\&&\null
 - 8 z \mp 16 x^4 z \pm 16 y^4 z + 8 x^4 + 16 x^2 y^2 + 8 y^4,
\\
d
&=&
4 (z^2 + 3 y^4 z + 4) x^2 z_x^2
 - 2 (z^2 + 12 x^2 y^2 z - 16) x y z_x z_y
\\&&\null
 + 4 (z^2 + 3 x^4 z + 4) y^2 z_y^2
 + 2 (z^2 + 8 x^2 y^2 z + 4 y^4 z + 20) x z z_x
\\&&\null
 + 2 (z^2 + 4 x^4 z + 8 x^2 y^2 z + 20) y z z_y
 + 4 (18 x^4 y^4 z^3 + z^3
\\&&\null
 + 3 x^4 z^2 - 2 x^2 y^2 z^2 + 3 y^4 z^2
   + 12 z + 4 x^4 + 8 x^2 y^2 + 4 y^4) z.
\end{eqnarray*}
The invariants $I_3,I_4,I_5$ are large fractions whose
non-reducible numerators are polynomials of order three in
$z_x,z_y$, five in $z$, and six in $x,y$. Invariants $I_s$, $s >
5$, are even more cumbersome.

Computation shows that the jacobian
$\partial(I_1,I_2,I_3,I_4,I_5)/\partial(x,y,z,z_x,z_y)$ is
nonzero, hence the first five invariants are functionally
independent and can be chosen to be local coordinates on
$J^1(\tau)$. Although an explicit inversion is rather hopeless,
one can still find algorithmically the relations connecting
principal invariants $I_1,\dots,I_5$ and  higher $I_k$ at least
in principle. This kind of procedure is outlined in
Example~\ref{Ex2} below.
\end{example}

\begin{example} \rm
\label{Ex2} Put $\zeta = z_x + z_y + e$ and consider the family of
equations
\begin{equation}
\label{eq2}
(4 z_x z_y + \zeta^2)(z_{xx} z_{yy} - z_{xy}^2)
 + 4 \zeta^2 (z_y z_{xx} + z_x z_{yy} + \zeta^2) = 0.
\end{equation}
depending on parameter $e$. Assuming that $e \not= 0$, we have
$$
\aligned
I_1 &= 2 \frac{(z_x + z_y)^2 + 3 e (z_x + z_y) + 4 e^2}
   {5 e z_x + e z_y + 4 e^2}, \\
I_2 &= 2 \frac{(z_x + z_y)^2 + 3 e (z_x + z_y) + 4 e^2}
   {e z_x + 5 e z_y + 4 e^2}, \\
I_3 &= 2^{3/2}\,
\frac{7 z_x^2 + 6 z_x z_y - z_y^2 + 33 e z_x + 5 e z_y + 21 e^2}
   {e^{1/2} (5 z_x + z_y + 4 e)^{3/2}}, \\
I_4 &= 2^{3/2}\,
\frac{-z_x^2 + 6 z_x z_y + 7 z_y^2 + 5 e z_x + 33 e z_y + 21 e^2}
   {e^{1/2} (5 z_x + z_y + 4 e)^{3/2}}, \\
I_5 &= 2^{5/2}\,
\frac{(z_x + z_y)^3 + 7 e (z_x + z_y)^2 + 17 e^2 (z_x + z_y) + 21 e^3}
   {e^{3/2} (5 z_x + z_y + 4 e)^{3/2}}.
\endaligned
$$
All invariants are independent of $x,y,z$, reflecting the fact that
$x \mapsto x + t_1$, $y \mapsto y + t_2$, $z \mapsto z + t_3$ are symmetries
of equation \eqref{eq2}.
One easily checks that $I_1,I_2$ are functionally independent, but it is
still not straightforward to express $z_x,z_y$ in terms of $I_1,I_2$ explicitly.

To establish the dependence of $I_s$, $s > 3$, on $I_1,I_2$, we
observe that for every $s$ there exists a polynomial
$P_s(z_x,z_y,I_s)$ such that $I_s$ is a solution of the equation
$P_s = 0$. Then what we need is eliminating $z_x,z_y$ from the
system
$$
\aligned
&I_1 - 2 \frac{(z_x + z_y)^2 + 3 e (z_x + z_y) + 4 e^2}
   {5 e z_x + e z_y + 4 e^2} = 0, \\
&I_2 - 2 \frac{(z_x + z_y)^2 + 3 e (z_x + z_y) + 4 e^2}
   {e z_x + 5 e z_y + 4 e^2} = 0, \\
&P_s(z_x,z_y,I_s) = 0.
\endaligned
$$
To this end, it suffices to compute the Gr\"obner basis of the
last system with respect to an ``elimination ordering'' of
monomials. With the help of the {\it Groebner} package of {\it
Maple 10} the following quadratic equation for $I_3$,
$$
\aligned
0 &= 4096\,I_2^6 I_3^2
 - I_2^3 (729\, I_1^3 I_2^3
 - 1971\, I_1^3 I_2^2
 + 20493\, I_1^2 I_2^3
\\&\qquad + 3563\, I_1^3 I_2
 - 51114\, I_1^2 I_2^2
 + 183915\, I_1 I_2^3
\\&\qquad  + 3951\, I_1^3
 - 52723\, I_1^2 I_2
 + 45517\, I_1 I_2^2
 + 102191\, I_2^3) I_3
\\&\quad + (27\, I_1^3 I_2^2
 - 81\, I_1^2 I_2^3
 - 32\, I_1^3 I_2
 + 426\, I_1^2 I_2^2
 - 1206\, I_1 I_2^3
\\&\qquad - 44\, I_1^3
 + 270\, I_1^2 I_2
 - 800\, I_1 I_2^2
 - 1114\, I_2^3)^2
\endaligned
$$
can be found rather quickly as well as similar quadratic equations
for $I_4,I_5$. The assumptions of Sect.~\ref{parallelism} are
satisfied as well. In particular, $\Delta_1,\Delta_2$ are nonzero
since
$$
\aligned
\Delta_1 = \Delta_2 = -128 &
\frac{(z_x + z_y) (3 z_x + 3 z_y + 8 e) (z_x + z_y + e)^4}
  {e^2 (z_x + 5 z_y + 4 e)^2 (5 z_x + z_y + 4 e)^2}
\\
&\times \frac{z_x^2 + 2 z_x z_y + z_y^2 + 3 e z_x + 3 e z_y + 4 e^2}
  {z_x^2 + 6 z_x z_y + z_y^2 + 2 e z_x + 2 e z_y + e^2}.
\endaligned
$$
This enables us to compute the higher invariants. For instance,
$I_6$ is solution of the quadratic equation
$$
\aligned
0 &= - 16 I_1^2
(27 I_1^4 I_2
 - 27 I_1^3 I_2^2
 + 22 I_1^4
 - 56 I_1^3 I_2
 - 2 I_1^2 I_2^2
\\&\qquad + 8 I_1^2 I_2
 - 42 I_1^3
 + 50 I_1 I_2^2
 + 28 I_1^2
 + 56 I_1 I_2
 + 28 I_2^2) I_6^2
\\&\quad
 + I_1 (I_1 I_2 - I_1 - I_2)
(9 I_1 I_2 + 7 I_1 + 7 I_2)
(3 I_1^3 I_2
 - 3 I_1^2 I_2^2
\\&\qquad - 26 I_1^3
 - 34 I_1^2 I_2
 - 8 I_1 I_2^2
 + 18 I_1^2
 + 36 I_1 I_2
 + 18 I_2^2) I_6
\\&\quad +
(I_1 + I_2)^2
(I_1 I_2 - I_1 - I_2)
(9 I_1 I_2 + 7 I_1 + 7 I_2)
(I_1 I_2 - 2 I_1 - 2 I_2)^2.
\endaligned
$$
Although every invariant computed so far depends on $e$, its
expression in terms of $I_1,I_2$ does not. This suggests the idea
that the parameter $e$ is removable. And indeed, after
substitution $z \mapsto e z$ equation \eqref{eq2} becomes
equivalent to itself with $e = 1$. Thus, the family of equations
\eqref{eq2} consists of a continuum of generic members with $e
\not= 0$, which are all mutually equivalent, and a single
non-generic member with $e = 0$ (in which case $\Lambda_1 = \Lambda_2 = 0$).
\end{example}

\begin{example} \rm
Consider the family of equations
$$
\textstyle\frac{1}{4} (z_{xx} z_{yy} - z_{xy}^2)
 + y^2 z_{xx} - 2 x y z_{xy} + x^2 z_{yy} + e x^2 y^2 = 0,
$$
depending on a real parameter $e \ne 4$.
Then the first five invariants are constants
$$
\begin{aligned}
&I_1 = I_2 = 2 \frac{e + 12}{e - 4},\\
&I_3 = I_4 = \frac{800}{e - 4}, \\
&I_5 = 3200 \frac{(e + 12)^2}{(e - 4)^3},
\end{aligned}
$$
while the higher invariants $I_s$ are undefined.

The equation belongs to the subclass ``h", or ``p'', or ``e"
(see~\ref{subtype}) if $e
> -4$ or $e = -4$ or $e < -4$, respectively.
\end{example}

\section*{Acknowledgements}

\noindent M. Marvan acknowledges the support from GA\v{C}R under
grant 201/04/0538.

%


\begin{thebibliography}{99}

\bibitem{ALV}
D.V. Alekseevskiy, A.M. Vinogradov and V.V. Lychagin,
Basic ideas and concepts of differential geometry.
in: {Geometry, I}
Encyclopaedia Math. Sci. 28,
Springer, Berlin, 1991, 1--264.

\bibitem{FN} A. Fr\"olicher and A. Nijenhuis, Theory of vector
     valued differential forms. Part I: Derivations in the graded
     ring of differential forms, {\it Indag. Math.} {\bf 18} (1956) 338--359.

\bibitem{H-W}
P. Hartman and A. Wintner, On hyperbolic partial differential equations,
{\it American Journal of Mathematics} {\bf 74} (1952) 834--864.

\bibitem{KLV} I.S. Krasil'shchik, V.V. Lychagin and A.M. Vinogradov,
     {\it Geometry of Jet Spaces and Nonlinear Partial Differential
     Equations}, Gordon and Breach, New York, 1986.

\bibitem{KV} I.S. Krasil'shchik and A.M. Vinogradov, Editors, {\it
  Symmetries and Conservation Laws for Differential Equations of
  Mathematical Physics}, Translations of Mathematical Monographs.
  Vol.182, Providence RI: American Mathematical Society, 1999.

\bibitem{Kr1}
B.S. Kruglikov,
Some classificational problems in four-dimensional geometry:
distributions, almost complex structures and the generalized Monge--Amp\`ere equations,
{\it Math. Sbornik} {\bf 189} (1998) (11) 61-74 (in Russian);
English translation in {\it Sb. Math.} {\bf 186} (1998) (11--12) 1643--1656;
e-print: http://xxx.lanl.gov/abs/dg-ga/9611005.

\bibitem{Kr2}
B.S. Kruglikov,
Symplectic and contact Lie algebras with application to the
Monge--Amp\`ere equations {\it Trudy Mat. Inst. Steklova} {\bf 221} (1998) 232---246 (in Russian); English translation in {\it Proc. Steklov Math. Inst.} {\bf 221} (1998) (2) 221--235;
e-print: http://xxx.lanl.gov/abs/dg-ga/9709004

\bibitem{Kr3}
B.S. Kruglikov,
Classification of Monge--Amp\`ere equations with two variables,
in: {\it Geometry and Topology of
   Caustics - CAUSTICS '98 (Warsaw)}, Banach Center Publications
   50 (Polish Acad. Sci., Warsaw, 1999) 179-194.
\bibitem{Ku}
A. Kushner, Monge--Amp\`ere equations and $e$-structures, {\it
Dokl. Akad. Nauk} {\bf 361} (1998) (5) 595--596.

\bibitem{Lewy} H. Lewy, \"Uber das Anfangswertproblem bei einer hyperbolischen
nichtlinearen partiellen Differentialgleichung zweiter Ordnung mit
zwei unabh\"angigen Ver\`anderlichen, {\it Math. Annalen} {\bf 98} (1928)
179--191.

\bibitem{Ly1} V.V. Lychagin, Contact geometry and non-linear second order
differential equations, {\it Russian Math. Surveys} {\bf 34} (1979) 149--180.

\bibitem{Ly2} V.V. Lychagin, {\it Lectures on Geometry of Differential
    Equations}, Universita ``La Sapienza,'' Roma, 1992, 133 p.

\bibitem{LRC} V.V. Lychagin, V.N. Rubtsov and I.V. Chekalov,
   {\it A classification of Monge-Ampere equations}, Ann. Sc.
   Ecole Norm. Sup. (4) 26 (1993), 281-308.

\bibitem{MVY}
M. Marvan, A.M. Vinogradov and V.A. Yumaguzhin,
     Differential invariants of generic hyperbolic Monge--Amp\`ere equations 
     {\it Russian Acad. Sci. Dokl. Math.} {\bf 405} (2005) 299--301 (in Russian)
     English translation in:
     {\it Doklady Mathematics} {\bf 72} (2005) 883--885.

\bibitem{Mats1}
M. Matsuda, Two methods of integrating Monge--Amp\`ere's equations,
{\it Trans. Amer. Math. Soc.} {\bf 150} (1970) 327--343.

\bibitem{Mats2}
M. Matsuda, Two methods of integrating Monge--Amp\`ere's equations. II,
{\it Trans. Amer. Math. Soc.} {\bf 166} (1972) 371--386.

\bibitem{Mori}
T. Morimoto, Monge--Amp\`ere equations viewed from contact geometry.
in: {\it Symplectic Singularities and Geometry of Gauge
   Fields} (Warsaw, 1995), 105--121, Banach Center Publ., 39, Polish Acad.
Sci., Warsaw, 1997.

\bibitem{Tch} O.P. Tchij, Contact geometry of hyperbolic
   Monge-Amp\`ere eqquations, {\it Lobachevskii Journal of
   Mathematics} {\bf 4} (1999) 109--162.

\bibitem{Tun}  D.V. Tunitskiy, Monge-Amp\`ere equations and functors of
characteristic connection,
{\it Izv. RAN, Ser. Math.} {\bf 65} (6) (2001) 173--222.

\bibitem{Tuni} D.V. Tunitskiy, On the global solvability of hyperbolic
Monge--Amp\`ere equations, {\it Izv. Ross. Akad. Nauk Ser.
   Mat.} {\bf 61} (1997), No. 5, 177--224 (in Russian);
translation in {\it Izv. Math} {\bf 61} (1997), No. 5, 1069--1111.

\bibitem{V2} A.M. Vinogradov, Geometry of nonlinear differential
     equations, manuscript, Diffiety Institute, 1987.

\bibitem{V1} A.M. Vinogradov, Scalar differential invariants,
      diffieties and characteristic classes, in: {\it Me\-cha\-nics,
      Analysis and Geometry: 200 Years after Lagrange,} ed.
      M. Francaviglia (North-Holland), pp.379--414,
      1991.

\bibitem{Vi-Yu} A.M. Vinogradov and V.A. Yumaguzhin,
Differential invariants of webs on 2-dimensional manifolds, {\it Mat.
          Zametki} {\bf 48} (1990), No. 1, 46--68 (in Russian).

\end{thebibliography}
\end{document}